\documentclass{webofc}
\usepackage[varg]{txfonts}   
\usepackage{lineno}
\begin{document}
\title{Measurement of the production of (anti)(hyper)nuclei}
%
%

\author{\firstname{Chiara} \lastname{Pinto}\inst{1}\fnsep\thanks{\email{chiara.pinto@cern.ch}} for the ALICE Collaboration 
}

\institute{Technische Universität München, Physik Department, James-Franck-Str. 1, 85748 Garching bei München
          }

\abstract{%
In recent years, ALICE has extensively studied the production of light (anti)(hyper)nuclei in different collision systems and center-of-mass energies. Nevertheless, the production mechanisms of light (hyper)nuclei is still unclear and under intense debate in the scientific community. Two classes of models are typically used to describe nuclear production: the statistical hadronisation model and the coalescence ones. In heavy-ion collisions, both models describe well the production yields of light nuclei and their ratios to the yields of hadrons, making it difficult to distinguish between the two. On the contrary, small collision systems, such as pp and p--Pb collisions, are ideal to study the (anti)(hyper)nuclei production mechanisms, thanks to the large separation between the different predictions. In this paper, recent results on light nuclei production measured with ALICE at the LHC in small collision systems are discussed in the context of the nucleosynthesis models, allowing us to exclude some configurations of the SHM and coalescence models in order to set tighter constraints to the available theoretical predictions. 
}
\maketitle
\section{Introduction}
\label{intro}

The production mechanism of light (anti)nuclei in high-energy hadronic collisions is under intense debate in the heavy-ion physics community, despite the large amount of measurements published in recent years.  
Currently, the experimental data are described by two classes of phenomenological approaches: the thermal-statistical and the coalescence models. 
In the Statistical Hadronization Model (SHM) \cite{SHM}, hadrons are produced by a thermally and chemically equilibrated source and their abundances are fixed at the chemical freeze-out. This model provides an excellent description of the measured hadron yields in central nucleus--nucleus collisions, corresponding to charged-particle multiplicities of the order of $10^3$ \cite{SHM_2}. For small systems, such as pp and p--Pb collisions, the production of light nuclei can be described using a different implementation of this model based on the canonical ensemble, where exact conservation of quantum numbers is required \cite{SHM_8}. Significant deviations in this case are observed between the experimental data and predictions from the canonical SHM \cite{hyp_pPb}. The mechanisms of hadron production and the propagation of loosely-bound states through the hadron gas phase are not addressed by this model. 

On the other hand, the production of light (anti)nuclei can be modelled via the coalescence of protons and neutrons that are close by in phase space at the kinetic freeze-out and match the spin, thus forming a nucleus \cite{coalescence}. The key observable of the coalescence models is the coalescence parameter $B_{\rm A}$, which is related to the production probability of a nucleus via this process and can be calculated from the overlap of the nucleus wave function and the phase space distribution of the constituents via the Wigner formalism \cite{coalescence_theory}. Experimentally, $B_A$ is obtained as the ratio between the invariant yield of a nucleus with mass number $A$ and the square of the yield of protons: 

\begin{equation}
B_{A} = { \biggl( \dfrac{1}{2 \pi p^{\mathrm A}_{\mathrm T}} \dfrac{ \mathrm{d}^2N_{\mathrm A}}{\mathrm{d}y\mathrm{d} p_{\mathrm T}^{\mathrm A}}  \biggr)}  \bigg/{  \biggl( \dfrac{1}{2 \pi p^{\rm p}_{\mathrm T}} \dfrac{\mathrm{d}^2N_{\mathrm p} }{\mathrm{d}y\mathrm{d}p_{\mathrm T}^{\mathrm p}} \biggr)^A},
\label{eq:BA}
\end{equation}

\noindent where the labels $A$ and p indicate the nucleus and the proton, respectively, and \mbox{$p_{\mathrm T}^{\rm p}$ = $p_{\mathrm T}^{\mathrm A}$/$A$}.

In this paper, recent results on light nuclei production are discussed in the context of the aforementioned nucleosynthesis models. 
In high multiplicity (HM) pp collisions, by combining the measurements of the production of (anti)nuclei and femtoscopic measurements, the coalescence parameters are compared with parameter-free coalescence predictions. In addition, by comparing the production of in-jet and out-of-jet (anti)deuterons, it is possible to observe an increase of the (anti)deuteron coalescence probability in jets as compared to that out of jets. Additional information can be extracted from the study of very large and extremely loosely bound objects such as the hypertriton ($^3_\Lambda$H). This nucleus has a large wave function, hence its production yield in small collisions is extremely sensitive to the nucleosynthesis models. 

\section{Testing coalescence models}
\label{sec-1}

\begin{figure}[h]
\centering
\includegraphics[width=0.44\textwidth]{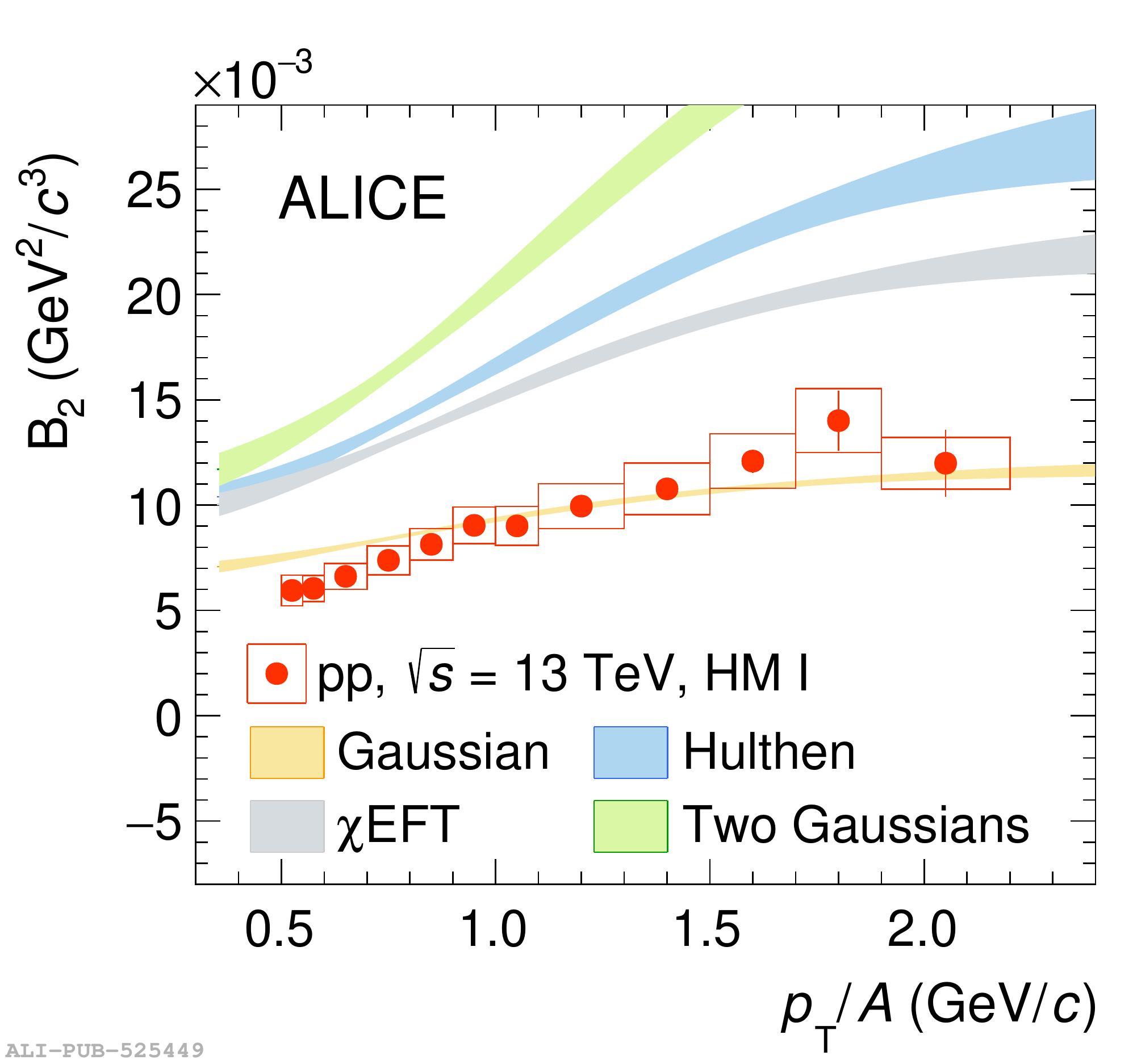}
\includegraphics[width=0.54\textwidth]{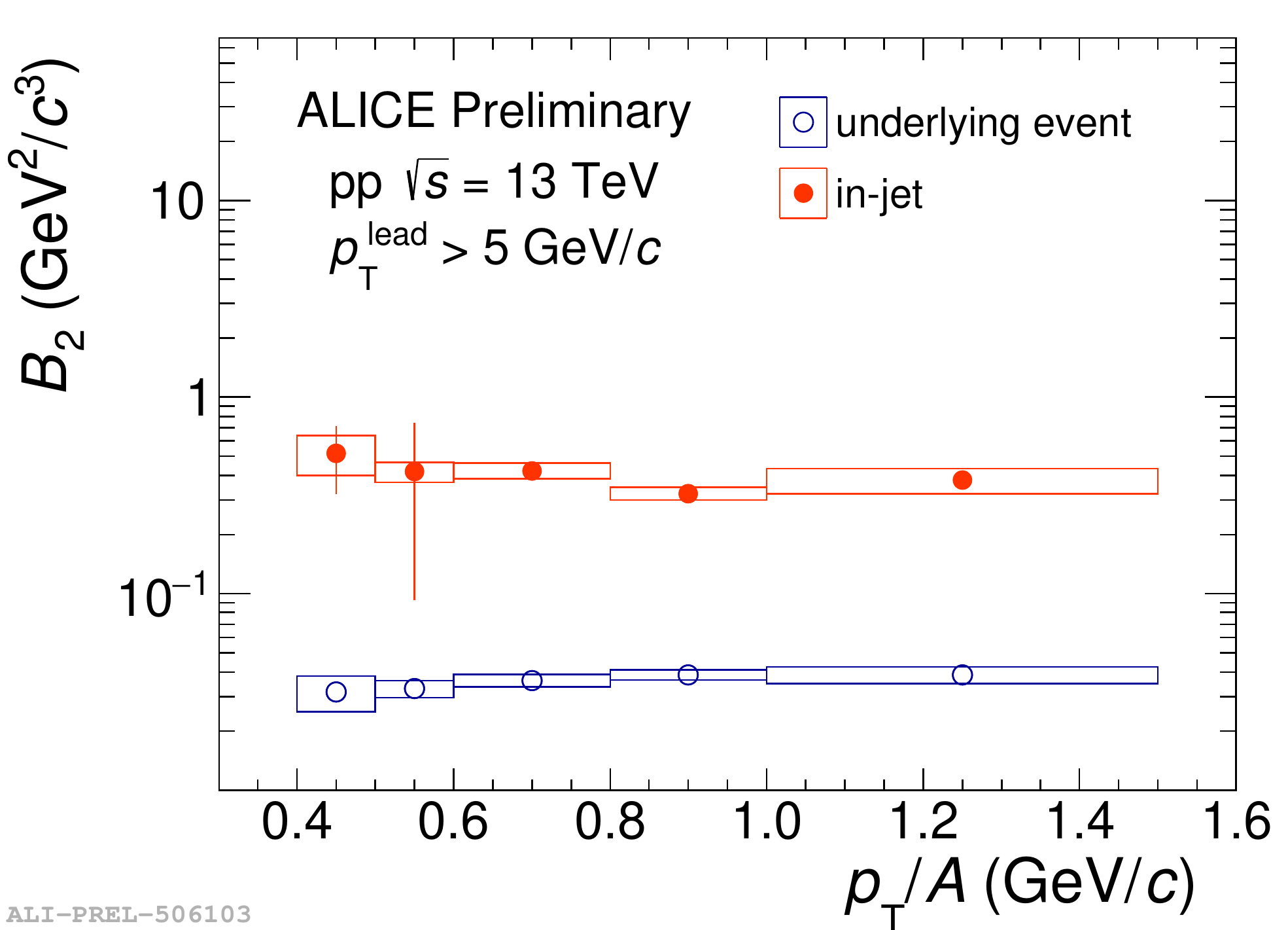}
\caption{\emph{Left.} Comparison between measurements and theoretical predictions for the coalescence parameter $B_2$ for (anti)deuterons as a function of $p_{\rm T}/A$. Theoretical predictions are obtained using different wave functions to describe nuclei: Gaussian (yellow), Hulthen (blue), $\chi$EFT (gray) and two Gaussians (green). \emph{Right.} Coalescence parameter $B_{2}$ in jets (full markers) and in the underlying event (empty markers) as a function of $p_{\rm T}/A$. }
\label{fig:B2HM}       
\end{figure}

In high multiplicity pp collisions at a center-of-mass energy $\sqrt{s}$ = 13 TeV, the coalescence models can be tested by combining the measurement of the production of light (anti)nuclei \cite{pp13TeVln} with the precise measurement of the source radius with femtoscopic techniques \cite{HMsource}. Specifically, it is possible to compare the experimentally measured coalescence parameter $B_2$ with parameter-free coalescence predictions \cite{coalB2}, where the only ingredients needed are the emission source size, which is measured for this specific dataset, and the deuteron wave function. As shown in the left panel of Fig. \ref{fig:B2HM}, several wave functions for deuterons have been tested. The Gaussian wave function provides the best description of the currently available ALICE data, despite the Hulthen one would be favoured by low-energy scattering experiments. These results show that additional studies in this direction are needed to further investigate the production mechanisms of nuclei. 

Further insights into the (anti)nuclei production mechanisms are obtained by measuring their production inside and outside of jets in pp collisions. Hadrons inside the jet cone are closer in phase space compared to hadrons outside of jets. In the coalescence picture, this condition should result in a larger coalescence probability in jets compared to that out of jets. The ALICE Collaboration has recently measured the deuteron production in jets and outside jets in pp collisions at $\sqrt{s}$ = 13 TeV. For this study, the leading particle with the highest transverse momentum ($p_{\rm T}^{\rm{ lead}} >$ 5 GeV/$c$) at midrapidity is used as a proxy for the presence of a jet. The (anti)deuteron production is measured in three different azimuthal regions, identified by the relative angular distance with respect to the leading particle. Three equal-size regions, $\pi/3$ wide, are consequently defined: the one around the leading particle (Toward), the one back-to-back to it (Away), and the one transverse to both of them (Transverse). The Toward and Away regions contain contributions from the leading and recoil jets in addition to the underlying event, while the Transverse region is dominated by the underlying event. Consequently, the transverse momentum distribution of deuterons in jets is obtained by subtracting the underlying event (Transverse contribution) from the Towards region. The results are found to be in agreement with those obtained with the two-particle correlation method \cite{jets}. 
The coalescence parameters $B_2$ in jets and in the underlying event are obtained as a function of the transverse momentum per nucleon ($p_{\rm T}/A$), using Eq. \ref{eq:BA}. As shown in the right panel of Fig. \ref{fig:B2HM}, the coalescence parameter in jets is a factor of about 15 larger than that in the underlying event. This observed enhancement is in agreement with the coalescence picture, and it can be interpreted as due to the reduced distance in phase space between nucleons in jets.

\section{Hypertriton measurement in small systems}
\label{sec-2}

\begin{figure}[h]
\centering
\includegraphics[width=0.5\textwidth]{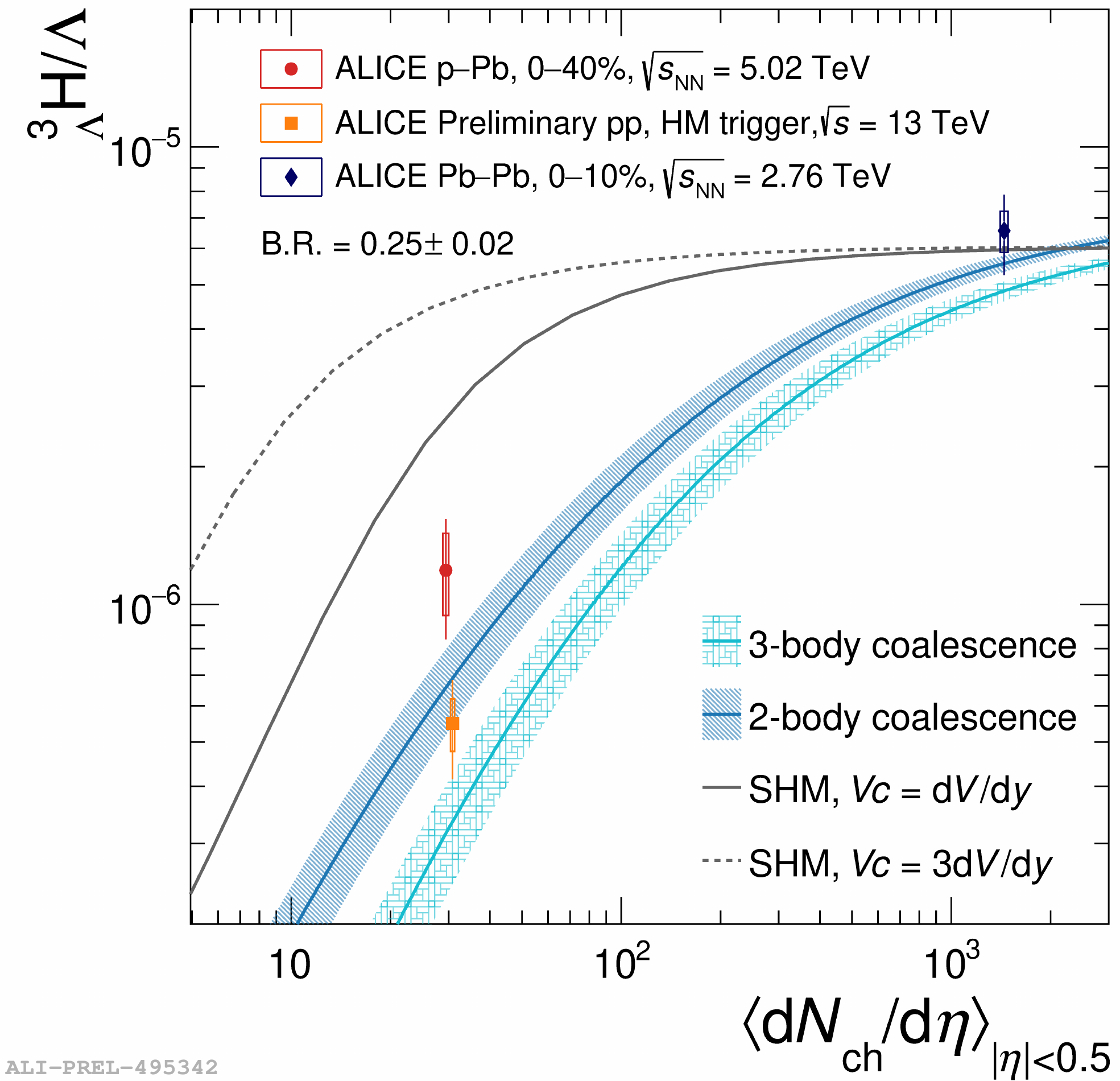}
\caption{Measurement of the $^3_\Lambda$H-to-$\Lambda$ ratio in pp, p--Pb \cite{hyp_pPb} and Pb--Pb collisions \cite{hyp_PbPb} as a function of the average charged-particle multiplicity density, in comparison with the expectations from the canonical SHM \cite{hyp_SHM} and the 2- and 3-body coalescence model \cite{hyp_coal}, shown as black lines, blue and light blue bands, respectively.}
\label{fig:HL}       
\end{figure}

A key measurement to understand the nuclear production mechanisms in high-energy collisions is the measurement of (anti)hypertritons in small systems \cite{hyp_pPb}. Hypertriton is a bound state of a proton, a neutron and a $\Lambda$ baryon. This state is characterised by a quite small $\Lambda$ separation energy, of the order of few hundreds of keV \cite{Lsepen}, and consequently by a wide wave function that can extend up to a radius of about 10 fm \cite{Lradius}. Therefore, the size of the $^3_\Lambda$H wave function is much larger than the hadron emission radius estimated with femtoscopic techniques in small systems (1--2 fm, \cite{ppradius, pPbradius}). For this reason, the $^3_\Lambda$H yield in pp and p--Pb collisions predicted by the coalescence models, where the ratio between the nucleus size and the source size directly influences its yield, is suppressed with respect to the statistical hadronisation model expectations, where the nuclear size does not enter explicitly. Hence, the measurement of the $^3_\Lambda$H-to-$\Lambda$ ratio in small systems is crucial to distinguish between the available nucleosynthesis models. 
In Fig. \ref{fig:HL}, the $^3_\Lambda$H-to-$\Lambda$ ratio measured in pp, p--Pb \cite{hyp_pPb} and Pb--Pb collisions \cite{hyp_PbPb} are compared with the expectations of the canonical statistical hadronization model \cite{hyp_SHM} and with the predicitions of the 2- and 3-body coalescence one \cite{hyp_coal}. The data show a good agreement with the 2-body coalescence predictions, while some tension with SHM at low charged-particle multiplicity density (d$N_{ch}$/d$\eta|_{|\eta|<0.5}$) is observed. In particular, the SHM configuration with a correlation volume equal to three units of rapidity $V_C$ = 3d$V$/d$y$ is excluded by more than 6$\sigma$ in the low d$N_{ch}$/d$\eta|_{|\eta|<0.5}$ region.

\end{document}